\RequirePackage[2020-02-02]{latexrelease}
\documentclass[twocolumn,tighten,times]{aastex631}
\usepackage{amssymb}
\usepackage{amsmath}
\usepackage{amsfonts}
\usepackage{hyperref}
\usepackage{float}
\usepackage{relsize}
\usepackage{verbatim}
\usepackage{color}
\usepackage{comment}
\usepackage{graphicx}
\usepackage{bm}
\usepackage{enumitem}
\usepackage[all]{hypcap} 

\makeatletter
\AtBeginDocument{%
    \catcode`_=12
    \begingroup\lccode`~=`_
    \lowercase{\endgroup\let~}\sb
    \mathcode`_="8000
    \immediate\write\@auxout{\catcode`_=12 }%
    \immediate\write\@auxout{\catcode`^=12 }%
}
\makeatother

\renewcommand{\ref}[1]{\hyperref[{#1}]{{\autoref*{#1}}}} 

\defcitealias{NG15}{NG15}

\newcommand{\todoblank}[1]{}

\graphicspath{{figures/}}

\def\be{\begin{equation}}
\def\ee{\end{equation}}
\def\ba{\begin{eqnarray}}
\def\ea{\end{eqnarray}}

\newcommand{\PSR}{PSR~J1455$-$3330}

\newcommand{\DM}{\ensuremath{\mathrm{DM}}} 

\newcommand{\Dtd}{\Delta t_{\rm d}}
\newcommand{\Dnud}{\Delta \nu_{\rm d}}

\renewcommand{\added}[1]{{#1}}
\renewcommand{\replaced}[2]{{#2}}
\renewcommand{\deleted}[1]{}

\shorttitle{Dispersion Measure Modeling for \PSR}
\shortauthors{Lam et al.}

\begin{document}

\title{
The NANOGrav 15-Year Data Set: A Case Study for Simplified Dispersion Measure Modeling for PSR J1455$-$3330 and the Impact on Gravitational Wave Sensitivity
}

\author[0000-0003-0721-651X]{Michael T. Lam}
\affiliation{SETI Institute, 339 N Bernardo Ave Suite 200, Mountain View, CA 94043, USA}
\affiliation{School of Physics and Astronomy, Rochester Institute of Technology, Rochester, NY 14623, USA}
\affiliation{Laboratory for Multiwavelength Astrophysics, Rochester Institute of Technology, Rochester, NY 14623, USA}
\email{}
\author[0000-0001-6295-2881]{David L. Kaplan}
\affiliation{Center for Gravitation, Cosmology and Astrophysics, Department of Physics and Astronomy, University of Wisconsin-Milwaukee,\\ P.O. Box 413, Milwaukee, WI 53201, USA}
\email{}
\author[0000-0001-5134-3925]{Gabriella Agazie}
\affiliation{Center for Gravitation, Cosmology and Astrophysics, Department of Physics and Astronomy, University of Wisconsin-Milwaukee,\\ P.O. Box 413, Milwaukee, WI 53201, USA}
\email{}
\author[0000-0002-8935-9882]{Akash Anumarlapudi}
\affiliation{Department of Physics and Astronomy, University of North Carolina, Chapel Hill, NC 27599, USA}
\email{}
\author[0000-0003-0638-3340]{Anne M. Archibald}
\affiliation{Newcastle University, NE1 7RU, UK}
\email{}
\author[0009-0008-6187-8753]{Zaven Arzoumanian}
\affiliation{X-Ray Astrophysics Laboratory, NASA Goddard Space Flight Center, Code 662, Greenbelt, MD 20771, USA}
\email{}
\author[0000-0003-2745-753X]{Paul T. Baker}
\affiliation{Department of Physics and Astronomy, Widener University, One University Place, Chester, PA 19013, USA}
\email{}
\author[0000-0003-3053-6538]{Paul R. Brook}
\affiliation{Institute for Gravitational Wave Astronomy and School of Physics and Astronomy, University of Birmingham, Edgbaston, Birmingham B15 2TT, UK}
\email{}
\author[0009-0002-1861-9787]{Olivia A. Combs}
\affiliation{Department of Physics, Lafayette College, Easton, PA 18042, USA}
\email{}
\author[0000-0002-6039-692X]{H. Thankful Cromartie}
\affiliation{National Research Council Research Associate, National Academy of Sciences, Washington, DC 20001, USA resident at Naval Research Laboratory, Washington, DC 20375, USA}
\email{}
\author[0000-0002-1529-5169]{Kathryn Crowter}
\affiliation{Department of Physics and Astronomy, University of British Columbia, 6224 Agricultural Road, Vancouver, BC V6T 1Z1, Canada}
\email{}
\author[0000-0002-2185-1790]{Megan E. DeCesar}
\altaffiliation{Resident at the Naval Research Laboratory}
\affiliation{Department of Physics and Astronomy, George Mason University, Fairfax, VA 22030, resident at the U.S. Naval Research Laboratory, Washington, DC 20375, USA}
\email{}
\author[0000-0002-6664-965X]{Paul B. Demorest}
\affiliation{National Radio Astronomy Observatory, 1003 Lopezville Rd., Socorro, NM 87801, USA}
\email{}
\author[0000-0001-8885-6388]{Timothy Dolch}
\affiliation{Department of Physics, Hillsdale College, 33 E. College Street, Hillsdale, MI 49242, USA}
\affiliation{Eureka Scientific, 2452 Delmer Street, Suite 100, Oakland, CA 94602-3017, USA}
\email{}
\author[0000-0001-7828-7708]{Elizabeth C. Ferrara}
\affiliation{Department of Astronomy, University of Maryland, College Park, MD 20742, USA}
\affiliation{Center for Research and Exploration in Space Science and Technology, NASA/GSFC, Greenbelt, MD 20771}
\affiliation{NASA Goddard Space Flight Center, Greenbelt, MD 20771, USA}
\email{}
\author[0000-0001-5645-5336]{William Fiore}
\affiliation{Department of Physics and Astronomy, University of British Columbia, 6224 Agricultural Road, Vancouver, BC V6T 1Z1, Canada}
\email{}
\author[0000-0001-8384-5049]{Emmanuel Fonseca}
\affiliation{Department of Physics and Astronomy, West Virginia University, P.O. Box 6315, Morgantown, WV 26506, USA}
\affiliation{Center for Gravitational Waves and Cosmology, West Virginia University, Chestnut Ridge Research Building, Morgantown, WV 26505, USA}
\email{}
\author[0000-0001-7624-4616]{Gabriel E. Freedman}
\affiliation{NASA Goddard Space Flight Center, Greenbelt, MD 20771, USA}
\email{}
\author[0000-0001-6166-9646]{Nate Garver-Daniels}
\affiliation{Department of Physics and Astronomy, West Virginia University, P.O. Box 6315, Morgantown, WV 26506, USA}
\affiliation{Center for Gravitational Waves and Cosmology, West Virginia University, Chestnut Ridge Research Building, Morgantown, WV 26505, USA}
\email{}
\author[0000-0001-8158-683X]{Peter A. Gentile}
\affiliation{Department of Physics and Astronomy, West Virginia University, P.O. Box 6315, Morgantown, WV 26506, USA}
\affiliation{Center for Gravitational Waves and Cosmology, West Virginia University, Chestnut Ridge Research Building, Morgantown, WV 26505, USA}
\email{}
\author[0000-0003-4090-9780]{Joseph Glaser}
\affiliation{Department of Physics and Astronomy, West Virginia University, P.O. Box 6315, Morgantown, WV 26506, USA}
\affiliation{Center for Gravitational Waves and Cosmology, West Virginia University, Chestnut Ridge Research Building, Morgantown, WV 26505, USA}
\email{}
\author[0000-0003-1884-348X]{Deborah C. Good}
\affiliation{Department of Physics and Astronomy, University of Montana, 32 Campus Drive, Missoula, MT 59812}
\email{}
\author[0000-0003-2742-3321]{Jeffrey S. Hazboun}
\affiliation{Department of Physics, Oregon State University, Corvallis, OR 97331, USA}
\email{}
\author[0000-0003-1082-2342]{Ross J. Jennings}
\altaffiliation{NANOGrav Physics Frontiers Center Postdoctoral Fellow}
\affiliation{Department of Physics and Astronomy, West Virginia University, P.O. Box 6315, Morgantown, WV 26506, USA}
\affiliation{Center for Gravitational Waves and Cosmology, West Virginia University, Chestnut Ridge Research Building, Morgantown, WV 26505, USA}
\email{}
\author[0000-0001-6607-3710]{Megan L. Jones}
\affiliation{Center for Gravitation, Cosmology and Astrophysics, Department of Physics and Astronomy, University of Wisconsin-Milwaukee,\\ P.O. Box 413, Milwaukee, WI 53201, USA}
\email{}
\author[0000-0002-0893-4073]{Matthew Kerr}
\affiliation{Space Science Division, Naval Research Laboratory, Washington, DC 20375-5352, USA}
\email{}
\author[0000-0003-1301-966X]{Duncan R. Lorimer}
\affiliation{Department of Physics and Astronomy, West Virginia University, P.O. Box 6315, Morgantown, WV 26506, USA}
\affiliation{Center for Gravitational Waves and Cosmology, West Virginia University, Chestnut Ridge Research Building, Morgantown, WV 26505, USA}
\email{}
\author[0000-0001-5373-5914]{Jing Luo}
\altaffiliation{Deceased}
\affiliation{Department of Astronomy \& Astrophysics, University of Toronto, 50 Saint George Street, Toronto, ON M5S 3H4, Canada}
\email{}
\author[0000-0001-5229-7430]{Ryan S. Lynch}
\affiliation{Green Bank Observatory, P.O. Box 2, Green Bank, WV 24944, USA}
\email{}
\author[0000-0001-5481-7559]{Alexander McEwen}
\affiliation{Center for Gravitation, Cosmology and Astrophysics, Department of Physics and Astronomy, University of Wisconsin-Milwaukee,\\ P.O. Box 413, Milwaukee, WI 53201, USA}
\email{}
\author[0000-0001-7697-7422]{Maura A. McLaughlin}
\affiliation{Department of Physics and Astronomy, West Virginia University, P.O. Box 6315, Morgantown, WV 26506, USA}
\affiliation{Center for Gravitational Waves and Cosmology, West Virginia University, Chestnut Ridge Research Building, Morgantown, WV 26505, USA}
\email{}
\author[0000-0002-4642-1260]{Natasha McMann}
\affiliation{Department of Physics and Astronomy, Vanderbilt University, 2301 Vanderbilt Place, Nashville, TN 37235, USA}
\email{}
\author[0000-0001-8845-1225]{Bradley W. Meyers}
\affiliation{Australian SKA Regional Centre (AusSRC), Curtin University, Bentley, WA 6102, Australia}
\affiliation{International Centre for Radio Astronomy Research (ICRAR), Curtin University, Bentley, WA 6102, Australia}
\email{}
\author[0000-0002-3616-5160]{Cherry Ng}
\affiliation{Dunlap Institute for Astronomy and Astrophysics, University of Toronto, 50 St. George St., Toronto, ON M5S 3H4, Canada}
\email{}
\author[0000-0002-6709-2566]{David J. Nice}
\affiliation{Department of Physics, Lafayette College, Easton, PA 18042, USA}
\email{}
\author[0000-0001-5465-2889]{Timothy T. Pennucci}
\affiliation{Institute of Physics and Astronomy, E\"{o}tv\"{o}s Lor\'{a}nd University, P\'{a}zm\'{a}ny P. s. 1/A, 1117 Budapest, Hungary}
\email{}
\author[0000-0002-8509-5947]{Benetge B. P. Perera}
\affiliation{Arecibo Observatory, HC3 Box 53995, Arecibo, PR 00612, USA}
\email{}
\author[0000-0002-8826-1285]{Nihan S. Pol}
\affiliation{Department of Physics, Texas Tech University, Box 41051, Lubbock, TX 79409, USA}
\email{}
\author[0000-0002-2074-4360]{Henri A. Radovan}
\affiliation{Department of Physics, University of Puerto Rico, Mayag\"{u}ez, PR 00681, USA}
\email{}
\author[0000-0001-5799-9714]{Scott M. Ransom}
\affiliation{National Radio Astronomy Observatory, 520 Edgemont Road, Charlottesville, VA 22903, USA}
\email{}
\author[0000-0002-5297-5278]{Paul S. Ray}
\affiliation{Space Science Division, Naval Research Laboratory, Washington, DC 20375-5352, USA}
\email{}
\author[0000-0003-4391-936X]{Ann Schmiedekamp}
\affiliation{Department of Physics, Penn State Abington, Abington, PA 19001, USA}
\email{}
\author[0000-0002-1283-2184]{Carl Schmiedekamp}
\affiliation{Department of Physics, Penn State Abington, Abington, PA 19001, USA}
\email{}
\author[0000-0002-7283-1124]{Brent J. Shapiro-Albert}
\affiliation{Department of Physics and Astronomy, West Virginia University, P.O. Box 6315, Morgantown, WV 26506, USA}
\affiliation{Center for Gravitational Waves and Cosmology, West Virginia University, Chestnut Ridge Research Building, Morgantown, WV 26505, USA}
\affiliation{Giant Army, 915A 17th Ave, Seattle WA 98122}
\email{}
\author[0000-0003-1407-6607]{Joseph Simon}
\altaffiliation{NSF Astronomy and Astrophysics Postdoctoral Fellow}
\affiliation{Department of Astrophysical and Planetary Sciences, University of Colorado, Boulder, CO 80309, USA}
\email{}
\author[0000-0001-9784-8670]{Ingrid H. Stairs}
\affiliation{Department of Physics and Astronomy, University of British Columbia, 6224 Agricultural Road, Vancouver, BC V6T 1Z1, Canada}
\email{}
\author[0000-0002-7261-594X]{Kevin Stovall}
\affiliation{National Radio Astronomy Observatory, 1003 Lopezville Rd., Socorro, NM 87801, USA}
\email{}
\author[0000-0002-2820-0931]{Abhimanyu Susobhanan}
\affiliation{Max-Planck-Institut f{\"u}r Gravitationsphysik (Albert-Einstein-Institut), Callinstra{\ss}e 38, D-30167 Hannover, Germany\\Leibniz Universit{\"a}t Hannover, D-30167 Hannover, Germany}
\email{}
\author[0000-0002-1075-3837]{Joseph K. Swiggum}
\altaffiliation{NANOGrav Physics Frontiers Center Postdoctoral Fellow}
\affiliation{Department of Physics, Lafayette College, Easton, PA 18042, USA}
\email{}
\author[0000-0001-9678-0299]{Haley M. Wahl}
\affiliation{Department of Physics and Astronomy, West Virginia University, P.O. Box 6315, Morgantown, WV 26506, USA}
\affiliation{Center for Gravitational Waves and Cosmology, West Virginia University, Chestnut Ridge Research Building, Morgantown, WV 26505, USA}
\email{}


\begin{abstract}

Evidence for a low-frequency gravitational-wave background using pulsar timing arrays has generated recent interest into its underlying contributing sources. However, multiple investigations have seen that the significance of the evidence does not change with choice of pulsar modeling techniques but the resulting parameters from the gravitational wave searches do. \PSR\ is one of the longest-observed pulsars in the array monitored by the North American Nanohertz Observatory for Gravitational Waves (NANOGrav) but showed no evidence for long-timescale red noise, either intrinsic or the common signal found among many pulsars in the array. In this work, we argue that NANOGrav's piecewise-constant function used to model variations in radio-frequency-dependent dispersive delay should not be used for this pulsar, and a much simpler physical model of a fixed solar wind density plus a linear \added{or quadratic} trend in dispersion measure is preferred. When the original model is replaced, (i) the pulsar's timing parallax signal changes from an upper limit to a significant detection, (ii) red noise becomes significant, and (iii) the red noise is consistent with the common signal found for the other pulsars. Neither of these signals are radio-frequency dependent. While the same physical motivation will not apply to many of the pulsars currently used in pulsar timing arrays, we argue for careful physically-motivated timing and noise modeling of pulsars used in precision timing experiments.

\end{abstract}

\keywords{pulsars: individual (\PSR) --- ISM: structure}

\section{Introduction}

The recent announcement of evidence for low-frequency gravitational waves (GWs) has provided not only another demonstration of the power of pulsar timing but also shown the power of correlating signals from many pulsars in a pulsar timing array (PTA) to uncover a common astrophysical signal \citep{NG15GWB,EPTA,PPTA}. With every Earth-pulsar line of sight (LOS) being unique, the different PTA collaborations needed to develop sophisticated timing models accounting for every revolution of each pulsar along with noise models that account for perturbations to the pulsar times of arrival (TOAs). The GW signal reported to date takes the form of a stochastic background, with increased variance at the longest timescales (decades) compared to the shortest cadences (weeks) observed.

The North American Nanohertz Observatory for Gravitational Waves (NANOGrav) has observed 68 millisecond pulsars (MSPs) to find evidence for correlations between the pulsars as a function of their angular separation on the sky. The observations and data are described in \citet{NG15}, along with the parameters used in the timing and noise modeling generally across the different pulsar data sets. An important time-variable effect to model is that of dispersion, a chromatic (radio-frequency-dependent) delay in the TOAs that scales with radio frequency $\nu$ as $\nu^{-2}$ as radio waves propagate through the intervening media. The time variations, if unmitigated, can lead to many microseconds of delay in the TOAs, far too large to allow detection low-Fourier-frequency GWs where hundreds of nanoseconds is required \citep{cs10}. In comparison to other collaborations, NANOGrav uses a piecewise-constant function (known as DMX; \citealt{NG5}) as part of the timing model to describe changes about a fiducial dispersion measure (DM), the integrated LOS electron density between the pulsar and Earth, whereas other collaborations have tended to describe changes as part of their noise model with a Gaussian process constrained with a power-law spectrum. NANOGrav's methodology has been to measure rapid changes in the DM which may not be picked up by the power-law spectrum descriptions alone. While such changes are known to exist, DMX has also been understood to absorb extra power in the overall fit \citep{Hazboun+2019} -- it is a conservative approach, though one that can lead to misestimation of the GW signal parameters of interest.

\begin{figure}[t]
\centering
\includegraphics[width=0.45\textwidth]{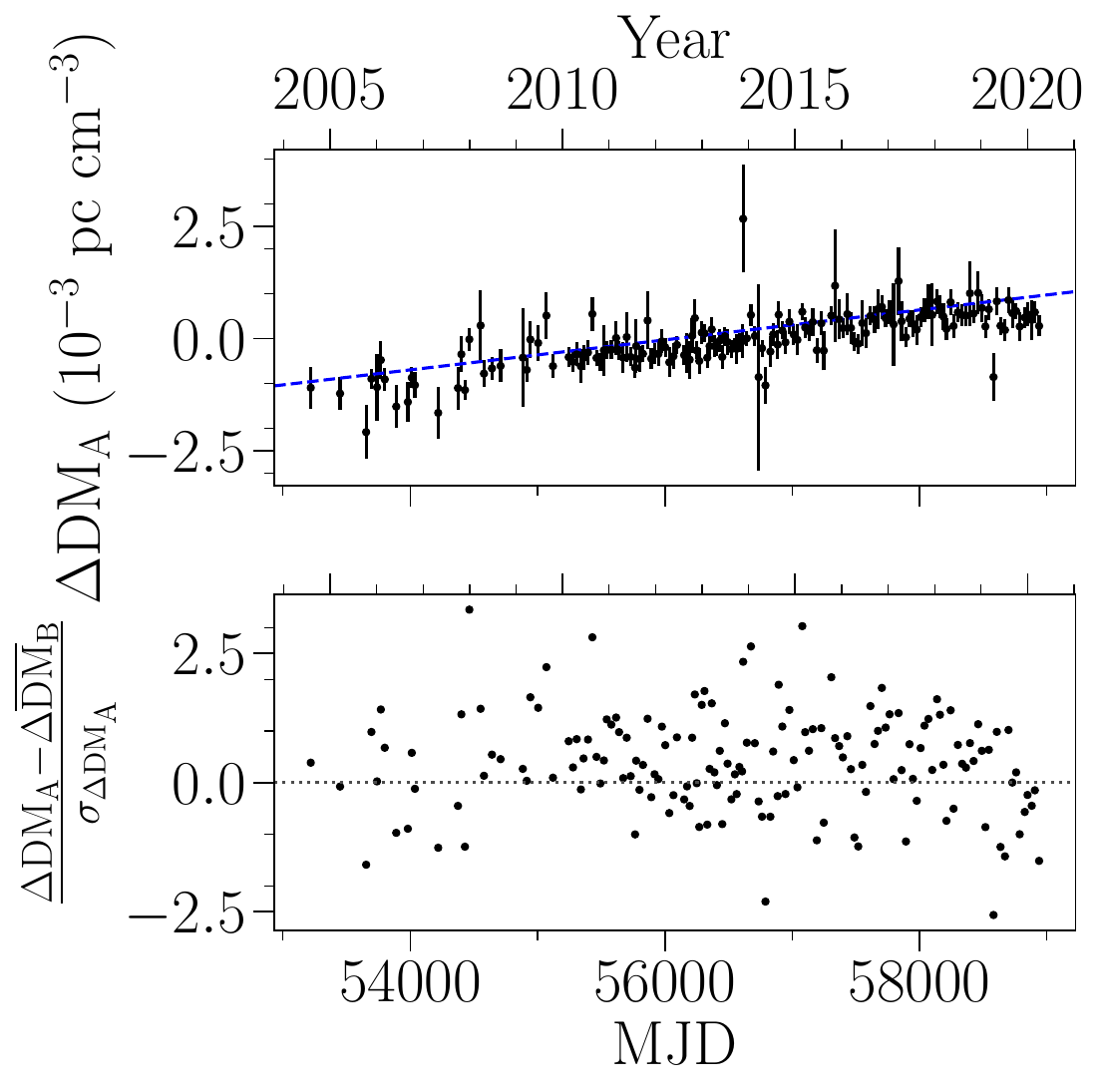}
\label{fig:DMA}
\caption{DMX timeseries (top) from the original 15-year data set, undergoing a single re-run using the \texttt{PINT} timing software, and modeled with DMX. We label this original model as Model A. The best-fit linear trend as fit from the multifrequency TOAs (model B, see \S\ref{sec:models}) is shown in blue. In the bottom, we subtract off this linear trend and divide by the uncertainties. All Model A DMX values are within 1.6$\sigma$ of the best-fit linear model.}
\end{figure}

\PSR\ is one of the longest-observed pulsars in the NANOGrav data set. It shows a DMX timeseries with a roughly linear trend, shown in Figure~\ref{fig:DMA}. Based on pulsar scintillation measurements, we argue that this trend \added{likely} does not come from the LOS crossing stochastic turbulent fluctuations in the density of the ionized interstellar medium (ISM). Indeed, \added{from a classical picture of turbulence} the predicted amplitude of these variations  based on the timespan of the observations is small in comparison to even the uncertainties on the measurements. \replaced{Therefore, modeling DM variations in the timing model with 157 independent parameters is ill-motivated from an Occam's Razor consideration for this specific pulsar.}{Therefore, as we do not expect significant DM variations from the turbulent fluctuations to be measurable on short timescales, modeling DM variations in the timing model with 157 independent parameters, as is done in the DMX model, appears unjustified for this specific pulsar from an Occam's Razor consideration.}

In this paper, we modify NANOGrav's standard timing analysis for \PSR\ with a set of alternative models. With different models, we find that {\it achromatic} timing model parameters, notably astrometric ones, can vary by a significant and peculiar amount. \added{We additionally find as a result that achromatic red noise is also absorbed in NANOGrav's standard timing analysis, which in turn removes timing perturbations consistent with the gravitational-wave background (GWB) signal we wish to find.} In conjunction with a long-track observation in which we measure the scintillation timescale for this pulsar, we argue that stochastic \added{DM} variations for the pulsar are small and the trend we see is dominated by a linear \added{or quadratic} component plus a small contribution from the LOS crossing the solar wind, providing additional evidence towards selecting one model in particular. In \S\ref{sec:DMt}, we describe the theoretical background for stochastic and systematic linear DM variations. In \S\ref{sec:obs}, we describe the timing and scintillation observations for \PSR\, and then in \S\ref{sec:predictedDMvar} we argue that the stochastic DM variations are small. In \S\ref{sec:models}, we implement multiple simplified timing models and discuss the results. Finally, in \S\ref{sec:implicationsDM} and \S\ref{sec:implicationsGW}, we discuss the implications for physical insights into the ISM and high-precision pulsar timing experiments, respectively, with the latter specifically aimed at GW detection and characterization. \added{We discuss flux density measurements for \PSR\ in Appendix~\ref{sec:appendix}}.

\section{Modeling Dispersion Measure Trends}
\label{sec:DMt}

Here we give a brief overview of stochastic trends due to turbulence in the ionized ISM as well as systematic linear trends due to LOS motions or density gradients.

\subsection{DM Variations from Turbulence}
\label{sec:turbulentDM}

As the Earth-pulsar LOS moves across the turbulent ISM, it traces through different electron-density fluctuations and thus the measured DM will vary over time. Assuming that the medium along the LOS is isotropic, then turbulence in the ISM electron density is well-described by a wavenumber spectrum that follows a power law over many orders of magnitude \citep{Armstrong+1995,Ocker+2021}, given by
\be 
P_{\delta n_e}(q, z) = C_n^2(z) q^{-\beta},\quad\quad q_{\rm outer} \leq q \leq q_{\rm inner}.
\label{eq:P_ne}
\ee
Here, $C_n^2$ is the amplitude of the wavenumber spectrum, $q$ is the wavenumber, $\beta$ is the spectral index, and $z$ is the position along the LOS. The outer and inner scales $l_{\rm outer}$ and $l_{\rm inner}$ relate to the outer and inner wavenumbers by $q_{\rm outer} = 2\pi/l_{\rm outer}$ and $q_{\rm inner} = 2\pi/l_{\rm inner}$, respectively. For a Kolmogorov medium, $\beta = 11/3$. The DM timeseries will have temporal variations also described by a power-law spectrum though with spectral index $\gamma = \beta - 1$, i.e., $P_\DM(f) \propto f^{-\gamma}$, where $f$ is the Fourier frequency and $\gamma = 8/3$ for a Kolmogorov medium.

One common method of quantifying the statistics of the turbulence in DM timeseries is via the DM structure function,
\be 
D_\DM(\tau) \equiv \langle\left[\DM(t+\tau) - \DM(t)\right]^2\rangle,
\label{eq:DMSF}
\ee
the average across times $t$ of the (squared) differences of DM measurements taken at a separation in time $\tau$. For an isotropic \deleted{Kolmogorov} medium given by the power-law spectrum above, the structure function can be shown to analytically evaluate to \citep{DMt}
\ba
D_\DM(\tau) & = & \frac{1}{(\lambda r_e)^2}\left(\frac{\tau}{\Dtd}\right)^{\alpha} \nonumber \\
& = & 1.47 \times 10^{-15}~(\mathrm{pc~cm^{-3}})^2 \left(\frac{\nu}{\mathrm{GHz}}\right)^2 \left(\frac{\tau}{\Dtd}\right)^{\alpha},
\label{eq:DMSFeval}
\ea
where \added{$\alpha = \beta - 2 = 5/3$ for a Kolmogorov medium}, $\lambda$ is the radio wavelength, $r_e$ is the classical electron radius, and $\Dtd$ is the scintillation timescale -- the timescale over which intensity maxima (``scintles'') in a dynamic spectra decorrelate. Eq.~\ref{eq:DMSFeval} then relates the short timescale $\Dtd$ with the amplitude of longer-term DM variations.

We can relate the structure function to the root-mean-square (rms) of the DM variations by simply
\be 
\sigma_\DM(\tau) = \left[\frac{1}{2} D_\DM(\tau)\right]^{1/2}.
\label{eq:sigmaDM}
\ee
Combining Eqs. \ref{eq:DMSFeval} and\, \ref{eq:sigmaDM} gives us a way to predict the expected DM variations from turbulent fluctuations over a timescale $\tau$ based on the observing radio frequency $\nu$ and the scintillation timescale $\Dtd$.

\subsection{Linear Variations in DM}
\label{sec:linearDM}

Many timeseries of DM show long-term linear trends. \citet{DMt} described several contributing factors from the changing LOS, including a change in the Earth-pulsar distance. The time derivative of DM relates to the pulsar and Earth motion parallel to the LOS,
\be 
\frac{d\DM}{dt} = n_e(\bm{x}_p) v_{p\parallel} - n_e(\bm{x}_e) v_{e \parallel} \approx n_e\left(v_{p \parallel} - v_{e \parallel}\right)
\label{eq:dDMdt}
\ee
where $\bm{x}_p$ is the pulsar position, $\bm{x}_e$ is the Earth position, and $v_{p\parallel}$ and $v_{e\parallel}$ are the pulsar and Earth speeds parallel to the LOS, respectively. \replaced{The second step approximates the entire medium with a uniform density, which}{The approximation in Eq.~\ref{eq:dDMdt} makes the assumption that the entire medium is uniform in density, and} evaluates to $10^{-5}~v_{100} n_{e,0.1}$~pc~cm$^{-3}$~yr$^{-1}$ for fiducial values of the relative velocity of 100 km~s$^{-1}$ \added{\citep{Matthews+2016}} and electron density of 0.1~cm$^{-3}$ \added{\citep{Draine}}. \added{Often $v_e < v_p$, sometimes substantially so, and therefore assuming $v_{p \parallel}$ is also much larger then the pulsar motion but also electron-density environment can regularly dominate \citep{DMt,Ocker+2024}.} We will show later that our preferred timing \replaced{model has}{models have} $d\DM/dt$ an order of magnitude higher than this value, which implies a larger but not unrealistic electron density at the pulsar if this parallel motion is the sole cause (a larger parallel velocity could contribute to this trend but likely not cause the entirety of it unless the pulsar had an unusually high velocity compared to the rest of the MSP population, see e.g., \citealt{Matthews+2016}). Transverse motion of the LOS across an electron-density gradient can also contribute to linear trends in DM, although refractive variations would be expected as well; \added{we found no evidence for refractive variations in the flux density measurements for \PSR, see Appendix~\ref{sec:appendix}.} \citet{DMt} also showed that fast changes in the slope of multiple linear trends could arise from ionized slabs, e.g., generated by the pulsar's relativistic particle wind. The implications of the measured linear trend from our modeling on the physical setup is discussed further in \S\ref{sec:implicationsDM}.

\subsection{DM Variations from the Solar Wind}

\label{sec:solarwind}

Free electrons expelled from the surface of the Sun form one component of the solar wind, and the changing Earth-pulsar LOS over the year causes an annual variation in pulsar DM timeseries, with especially large amplitudes for pulsars closer to the ecliptic plane.  Additional variations result depending on the solar latitude from which the particles are ejected, the solar cycle, coronal mass ejections, etc. Different models of the solar wind have been used in pulsar timing, for example, assuming that the electron density is radially symmetric and fixed over time \citep[e.g.,][]{Madison+2019}, assuming variations over time \citep[e.g.,][]{Tiburzi+2021}, using prior time- and solar-latitude-dependent information from other observatories \citep[e.g.,][]{You+2007}, or using Gaussian processes with fixed spectral properties \citep[e.g.,][]{Hazboun+2022}. Time independence is not well justified, with variations in electron density seen across very short timescales (including within a pulsar observation, e.g., \citealt{Kumar+2022}) and even across the 11-year solar cycles. The radial dependence of the electron density far from the Sun often follows a $1/r^2$ form with distance $r$ from the Sun, but closer in may not; see \citealt{Hazboun+2022} for a thorough overview of these forms. Typical electron densities at 1 AU are in the $\sim$2--10~cm$^{-3}$ range. For reference, using a fixed solar wind density of 7.9~cm$^{-3}$ \citep{Madison+2019} and the pulsar's ecliptic latitude of $-16.045^\circ$, the DM contribution at minimum solar elongation \citep{Hazboun+2022} is $4.0 \times 10^{-4}$~pc~cm$^{-3}$. \added{Using the fiducial value of solar wind density of 4.0~cm$^{-3}$ from the \texttt{tempo2} software package \citep{tempo2}, which is consistent with older spacecraft \citep{Issautier+1998,Issautier+2001} and pulsar timing \citep{Splaver+2005} measurements but also broadly consistent with recent estimates at these ecliptic latitudes \citep{Susarla+2024}, halves the contribution to $2.0 \times 10^{-4}$~pc~cm$^{-3}$.} \replaced{This is the amplitude}{These values provide an approximate range} of DM variations we expect each year due to the solar wind.


\section{Observations of \PSR}

\label{sec:obs}

The data used in this work are part of the NANOGrav 15-yr data set, with pulse profiles that are reduced and processed and from which TOAs are computed. \PSR\ was observed approximately monthly by the Green Bank Telescope of the Green Bank Observatory with both the 820~MHz and L-band (\replaced{1400}{1500}~MHz) receivers to estimate the DM delay. Earlier data were recorded with the lower bandwidth (64 MHz) GASP backend, while in 2010 we transitioned to using the GUPPI backend with 200 and 800 MHz of bandwidth covering each receiver, respectively. \added{After polarization and flux calibration and processing, GASP data were channelized to 4 MHz sub-bands while the GUPPI channelization yielded 1.5625 MHz sub-bands. For GUPPI, the TOAs from which DMs were then derived were generated from profiles further averaged so that the channelization was 12.5 MHz for L-band and 3.125 MHz for 820~MHz.} The typical observation length is $\sim$25 minutes, although included in the data set are data from longer tracks on the source, discussed in the section below, with no other observational setup changes.

The calibration and data reduction routines are described in further detail in \citet[][hereafter \citetalias{NG15}]{NG15}. In short, the profiles underwent basic radio frequency interference (RFI) excision for known sources and then with a per-subintegration algorithm that checks the off-pulse intensity variation in a rolling 20-frequency-channel-wide window. Full polarization information was recorded, and we observed a broadband noise source prior to each observation to calibrate differential gain and phase offsets between the two hands of polarization, and this noise source was calibrated against the bright unpolarized quasar B1442+101 monthly for absolute calibration. For this work, we averaged the polarizations to form total intensity profiles, and we fully time-averaged the profiles to compute TOAs per radio-frequency channel. More information about the RFI excision and polarization calibration can be found in \citet{NG9}, and the overall details of the data reduction in \citetalias{NG15}.

In the end, we analyzed 10818 TOAs. The timing models we analyzed in \S\ref{sec:models} have the same 13 radio-frequency-independent parameters modeling spin, astrometric, and binary terms, as well as two frequency-dependent parameters describing the profile evolution across the bands. In performing the timing fits for each of our models, we followed an approach similar to \citetalias{NG15}. We started with the 15-year data set parameter file as our initial estimate of the timing parameters, used the Bayesian pulsar timing analysis code \texttt{enterprise}  \citep{enterprise} to estimate the white and red noise parameters, and then fixed the maximum a posteriori noise parameters while refitting the timing model to arrive at our final timing solution.

\section{Predicted DM Variations for \PSR\ from the Dynamic Spectrum}
\label{sec:predictedDMvar}

\begin{figure}[t]
\hspace{-0.2in}
\includegraphics[width=0.5\textwidth]{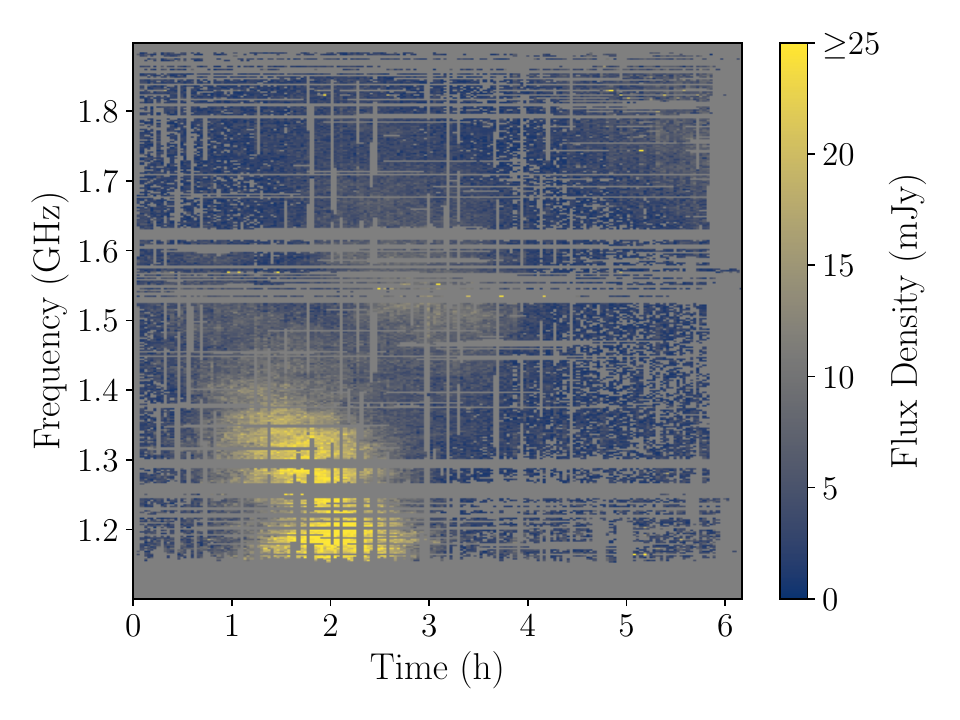}
\label{fig:dynspec}
\caption{Dynamic spectrum for MJD 56645 as measured by \texttt{PyPulse}. Brighter (yellower) patches denote increased flux density. The distorted gray lines/groupings denote zero-weighted pulses due to RFI flagging. \added{Remaining RFI causes several bright pixels to be seen, and to reduce the dynamic range all values above 25 mJy are displayed with the brightest yellow color. }}
\end{figure}

For most MSPs timed by NANOGrav we cannot obtain reliable estimates of the scintillation timescales as they are of the order of the length of a typical observation. Multi-hour tracks on pulsars can often help us estimate this parameter for individual sources \citep{longtracks}. As part of an observing campaign\footnote{GBT 13A-446, PI T.\ Pennucci} to estimate Shapiro delay signatures from binary companions in TOAs, on MJD 56645 we observed \PSR\ for 6.2 hours. The dynamic spectrum $I(t, \nu)$ as calculated by \texttt{PyPulse} \citep{pypulse} for this observation is shown in Fig.~\ref{fig:dynspec}. \texttt{PyPulse} computes the dynamic spectrum by fitting the known pulse template from \citetalias{NG15} to each $I(t, \nu)$ profile. The best-fit amplitude for each is saved to create the dynamic spectrum. A zero is saved for profiles in which RFI was excised. An earlier long-track observation on MJD 56563 was less than two hours long, and we were unable to extract the scintillation timescale and so ignored the observation further.

We can estimate the scintillation timescale by first calculating the 2D autocorrelation functon (ACF) of the dynamic spectrum, $R_I(\delta t, \delta \nu) = \langle I(t, \nu) I(t + \delta t, \nu + \delta \nu)\rangle$, where $\delta t$ and $\delta \nu$ are shifts in time and radio frequency, respectively. \added{Given the large bandwidth over which the scintles will vary drastically in size, we stretched the dynamic spectrum as in \citet{Levin+2016} and \citet{Turner+2021} prior to computing the ACF so that scintiles would have comparable width in the computation. We referenced our measurements to the band center frequency of 1500 MHz and assumed a Kolmogorov frequency scaling\footnote{\citet{Turner+2021} found with simulations that the wrong assumed index resulted in fractional errors $\approx$ 10\%. When we do not stretch the dynamic spectrum first here, we find $\Dtd = 4500 \pm 1100$~s, a 10\% change, but $\Dnud = 140 \pm 50$, a 33\% change but still within the uncertainties of the stretched measurements.
} of $\Dnud \propto \nu^{4.4}$. } Since we do not have a large number of scintles, we opted to estimate the scintillation parameters using 1D slices of the 2D ACF rather than fitting a full Gaussian to the central feature. The 1D slice along $R_I(\delta t, \delta \nu = 0)$ was fit with a Gaussian function, and the scintillation timescale was estimated with the typically-used half-width at the $1/e$ height \citep{Cordes2002}. Referenced to the center frequency of 1500~MHz, we find $\Dtd = 5000 \pm 1400$~s, where the uncertainty is dominated by finite-scintle errors \citep{Cordes1986}. The 1D slice along the other direction, along $R_I(\delta t=0, \delta \nu)$ was fit with a Gaussian for completeness, with the typically used half-width at half maximum. Again referenced to 1500~MHz and including finite-scintle errors, we measure $\Dnud = 210 \pm 80$~MHz. \added{Both measured values match what we see visually with the bright scintle in the bottom left of Figure~\ref{fig:dynspec} (when scaled to 1500 MHz) and also the fainter patch in the center.}

\added{
\subsection{How Representative are These Scintillation Measurements?}

To test how representative this epoch is in comparison to others, we performed two separate analyses: one on dynamic spectra generated using the method above on the available NANOGrav 15-year data set 1500~MHz data and one on flux density measurements. The dynamic spectra analysis is presented below. We see in Appendix~\ref{sec:appendix} that the flux density values are entirely consistent with the averaged values at other epochs and so do not comment on them further here.

We computed the ACFs in the same procedure as above but were unable to measure scintillation timescales owing to the $\sim$30 minute observing times. Instead, we estimated scintillation bandwidths and show those in Figure~\ref{fig:dnud}. We were unable to estimate $\Dnud$ for all epochs due to, for example, low pulse signal-to-noise ratio (S/N) or significant signal removed because of RFI. For many of our ACFs, there remains significant variation at very large lag, and so our Gaussian fits can be sensitive to which values we fit, and sometimes cannot find the central ACF feature. Arguments provided to \texttt{PyPulse} restrict the range to help the least-squares algorithm localize this feature, and for most epochs we set the window size to $\pm 200$ channels (recall that each channel has the same frequency since they are all stretched and then referenced to the same frequency, so one channel is approximately 1.5625 MHz for features in the center of the dynamic spectrum) but on MJDs 56790, 58848, and 58881 we had to restrict the range to $\pm 150$ channels based on visual inspection of the fits. On MJDs 55429, 56645 (the long track epoch), 58187, and 58545, we used the default window size equal to the number of stretched channel (685), which are shown as the four larger values with larger uncertainties in Figure~\ref{fig:dnud}. For such large scintles, the ACFs do not reach zero at large lags and so our estimates are somewhat sensitive to the choice in this window size but still within the displayed uncertainties; put another way, our displayed uncertainties for these measurements are underestimated (for the long-track scintillation timescale, the ACF does reach zero at large lags and so is minimally affected). We note that from visual inspection of the dynamic spectra, we do see larger than usual scintles for these epochs, and so while these four epochs are still constistent with the other measurements based on the uncertainties, the cause for the discrepancy is unclear. \citet{Coles+2010} show spiky features in scintillation bandwidth timeseries based on simulations and point to issues with the method of estimation from the ACF, but also show some statistical spread in individual scintle size in their simulation even when generated (see their Figure 1), which is a possible cause here but we cannot confirm with our data. They find that spikes in the scintillation timescale do not correlate with bandwidth, whereas both quantities should be correlated for longer-timescale refraction through a single plasma screen \citep{cpl86}. In summary, our long-track scintillation bandwidth measurement is loosely consistent with the typical values though the scintles are visually larger for this epoch along with a few others, and we cannot make definitive claims about the behavior of our scintillation timescale measurement.

\begin{figure}[t]
\includegraphics[width=0.5\textwidth]{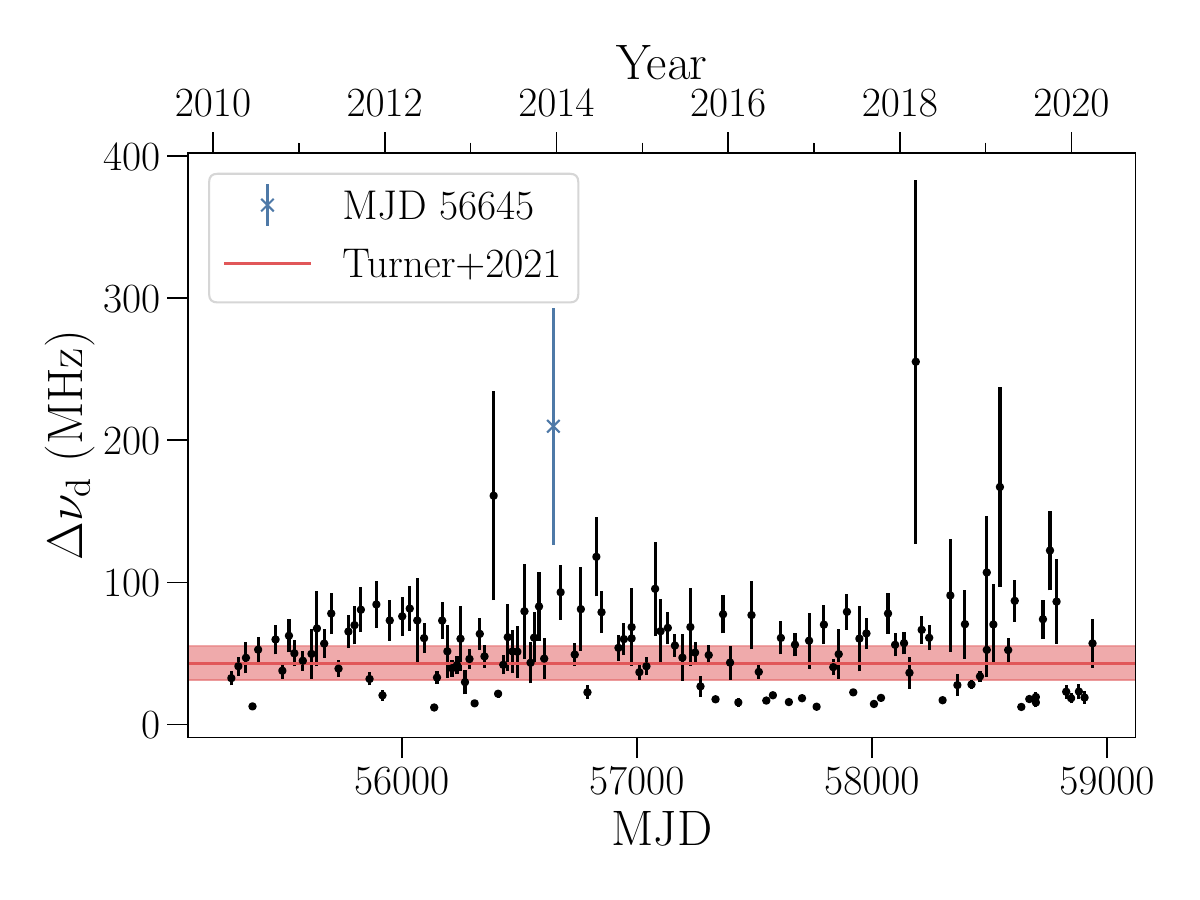}
\label{fig:dnud}
\caption{Scintillation bandwidths for the GUPPI 1500 MHz data from the NANOGrav 15-year data set. The blue cross is the measurement from the long track. The red band shows the weighted mean value reported from previous analyses of the NANOGrav 12.5-year data set \citep{Turner+2021}, $\Dnud = 43 \pm 12$ MHz. \citet{Levin+2016} found a slightly higher $\Dnud = 70 \pm 18$ MHz from analyses of the NANOGrav 9-year data set. Both of those works used the same methods which were similar in spirit to those here but with some processing and fitting differences. We see increases in the uncertainties as the values increase, largely dominated by the finite scintle effect \citep{Cordes1986}.
 }
\end{figure}

\subsection{Implications of the Measured Scintillation Timescale for DM Variations}
\label{sec:heuristic}
}

We adopt a heuristic argument to estimate the predicted variations we expect to observe in our DM timeseries. Consider the rms over the entire 15.7-yr timespan of the data, i.e., $\tau = 15.7$~yr. At a \replaced{center frequency of 1500~MHz}{representative middle frequency across our two bands of 1300~MHz}, we can use Eq.~\ref{eq:DMSFeval} to estimate the DM structure function followed by Eq.~\ref{eq:sigmaDM} to estimate the rms DM. \added{Assuming a Kolmogorov index $\alpha = 5/3$ and} using the nominal 5000~s value we find for $\Dtd$ above, we obtain $\sigma_\DM(15.7~\mathrm{yr}) = 5.1 \times 10^{-4}$~pc~cm$^{-3}$, \deleted{just over} twice the median uncertainty on the DM values found in the NANOGrav 15-yr data set for \PSR\ ($2.5 \times 10^{-4}$~pc~cm$^{-3}$, mean uncertainty of $3.2 \times 10^{-4}$~pc~cm$^{-3}$). That is, the expected variation in DM due to turbulence over the {\it entire} 15.7 years is of order the DM uncertainties on a single measurement. Variations across the timescale of $\tau = 1$~yr fall well below the uncertainty, with $\sigma_\DM(1~\mathrm{yr}) = 0.6 \times 10^{-4}$~pc~cm$^{-3}$. From \S\ref{sec:solarwind}, these uncertainties are also of order the maximum amplitude of DM variations from the change in solar elongation. \added{Therefore, we do not expect DMX to measure any significant variations from either of these effects, especially on the timescale of our monthly cadence.}
\replaced{And so while the DMX model may still be used to find rapid variations in DM such as from plasma lensing or even solar flares, for variations due to the turbulent spectrum in Eq.~\ref{eq:P_ne}, many more parameters are being fit than are required from an Occam's Razor perspective.}{In general, the DMX model may still be useful in finding rapid variation in DM, such as from plasma lensing or even solar flares, even in this pulsar. For variations in this pulsar's DM due to the turbulent spectrum in Eq.~\ref{eq:P_ne}, many more parameters are being fit than are justified from an Occam's Razor perspective.}

\added{
The heuristic argument above assumes a Kolmogorov medium along the line of sight and implies that stochastic variations are small compared with the observed trend and are of order the DM uncertainty. To explore this implication empirically, we calculated the DM structure function for the DMX timeseries, shown in Figure~\ref{fig:DMSF}. We show the extrapolation (purple) of the scintillation timescale measurement (black circle) assuming a Kolmogorov medium and see that it falls well below the measured structure function. In log-log space, we fit a power-law plus noise model to the structure function, $D_{\rm DM}(\tau) = C\tau^\alpha + 2\sigma_{\rm n}^2$, and the estimated $\alpha = 2.0 \pm 0.1$, which is consistent with a linear trend but also a square-law turbulent medium \citep{DMt,sc17} over our range of electron-density wavenumbers. If we fix the constant $C$ using our measured $\Dtd$ according to Eq.~\ref{eq:DMSFeval} and assuming a middle frequency of 1300~MHz, the best-fit estimate $\alpha = 1.822 \pm 0.003$ and we show the extrapolation (blue) from the scintillation timescale assuming this index in Figure~\ref{fig:DMSF} as well. Note that the fit for $C$ yields $1.0 \pm 0.9 \times 10^{-13}\ (\mathrm{pc}~\mathrm{cm}^{-3})^2$, which dominates the brown uncertainty region in the figure and both the spread and the fitting error on $\alpha$ are reduced substantially. Also note that any true deterministic linear trends will bias fits to the structure function \citep{DMt,NG9DM}, and so we caution against making inferences on electron-density wavenumber spectral index $\beta$ from the structure function itself or assuming the quoted stochastic uncertainty provides an accurate estimate of the overall uncertainty. However, we cannot rule out that the trend we are seeing is a slowly-varying linear segment a stochastic process representing a steeper-than-Kolmogorov turbulent medium; further DM measurements will help differentiate between the two possibilities. Regardless, in either case the DM structure function shows empirical evidence consistent with the linear trend seen in the DMX timeseries, and a reduced number of DM parameters is still warranted.

\begin{figure}[t]
\includegraphics[width=0.5\textwidth]{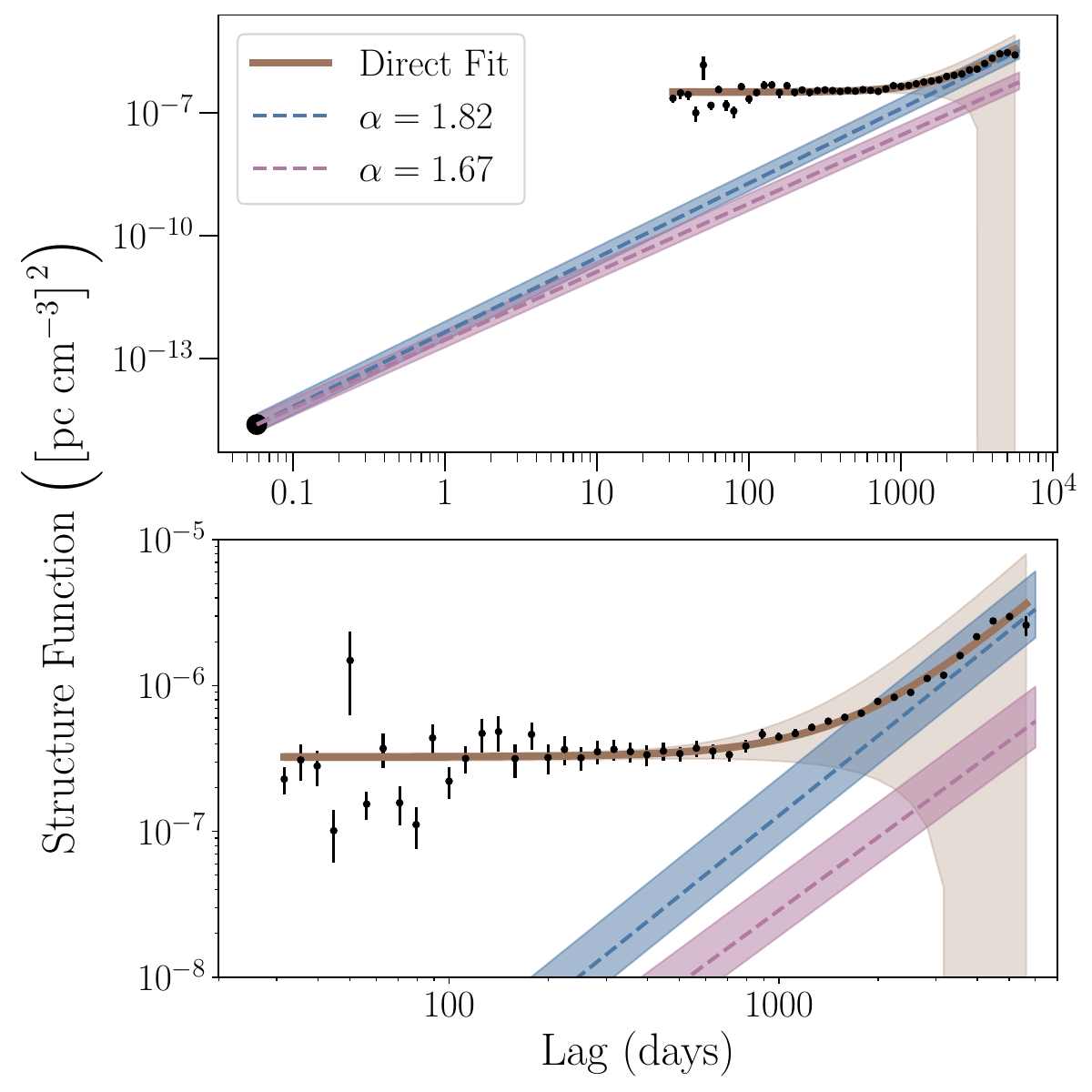}
\label{fig:DMSF}
\caption{DM structure function (black points) calculated from the DMX timeseries. We performed standard uncertainty propagation and assumed no correlations or realization errors \citep{NG9DM}. The bottom panel is a zoom in of the top panel. We directly fit a power-law plus noise, $D_{\rm DM}(\tau) = C \tau^\alpha + 2 \sigma_n^2$ \citep{DMt}, and show the best fit model with uncertainties propagated (brown line and shaded region). The larger black circle is the value of the DM structure function at the measured scintillation timescale $\Dtd = 5000$~s. A power-law extrapolation assuming a Kolmogorov medium is shown in purple. Fixing $C$ based on Eq.~\ref{eq:DMSFeval} and assuming a middle frequency of 1300~MHz yields $\alpha = 1.82$; we show the power-law extrapolation assuming this $\alpha$ value in blue. The shaded regions for these extrapolations come from the uncertainty on $\Dtd$ of 1400~s. True linear trends will bias fits to the structure function and we caution against making inferences on $\beta$ as a result.
 }
\end{figure}
}

\section{Simplified Timing Models for \PSR}
\label{sec:models}

Given the arguments that stochastic DM variations from turbulence are small over our timespan as laid out in \S\ref{sec:predictedDMvar}, we fit several simplified DM models. The DM models are described below and the parameter estimates and likelihood values resulting from refitting the timing models are provided in Table~\ref{table:J1455}. We used the \texttt{PINT} software package \citep{PINT} to evaluate the best-fit parameters for each timing model \added{and used \texttt{pint\_pal} \citep{pintpal} with underlying calls \texttt{enterprise}  \citep{enterprise} to perform the default noise modeling with 100000 Markov Chain Monte Carlo steps. Following our standard procedures \citep{NG15}, we reran the timing-model fit in \texttt{PINT} after the noise modeling and report those parameters.}.

\begin{figure}[t!]
\centering
\includegraphics[width=0.48\textwidth]{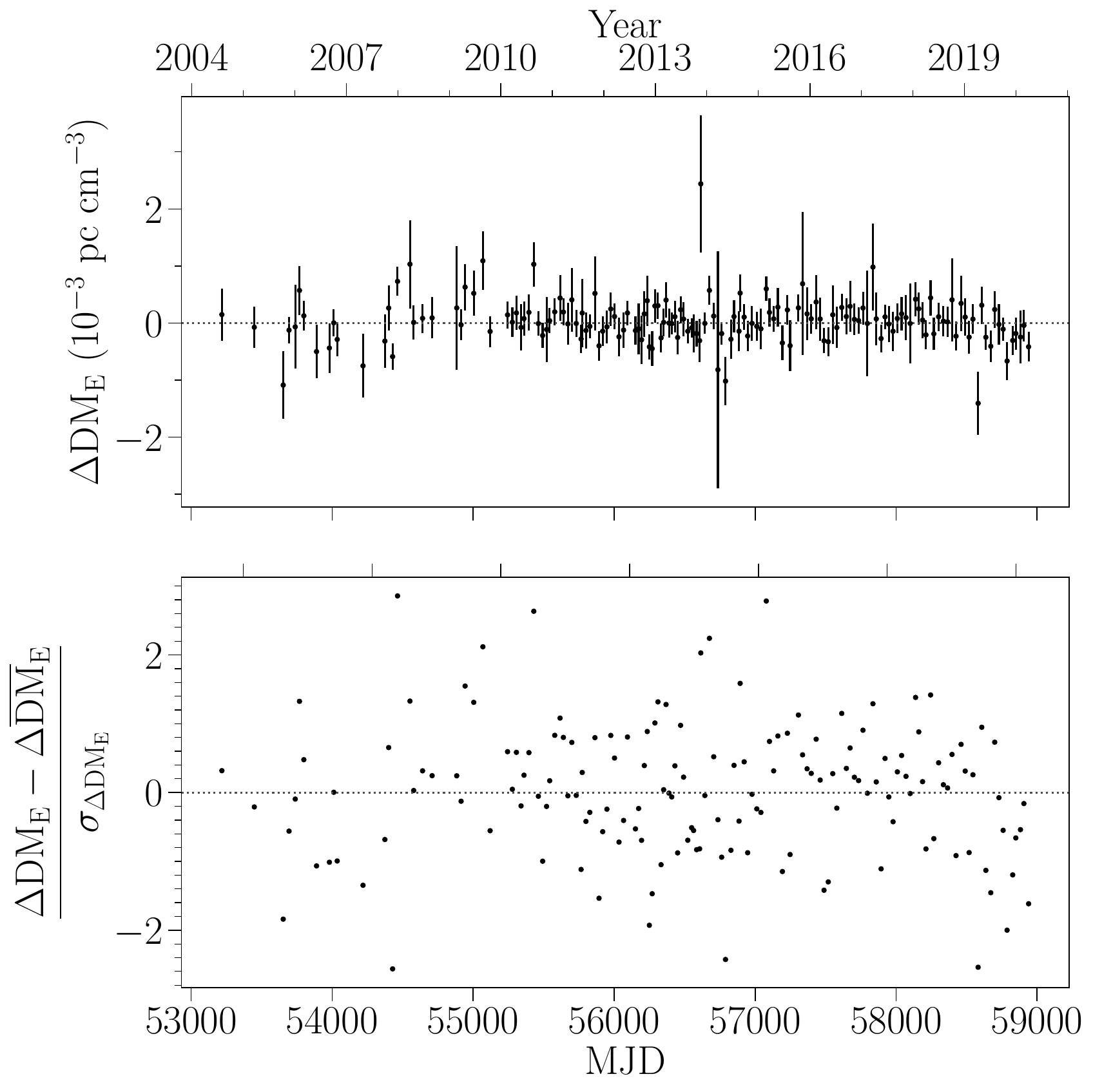}
\label{fig:ModelE}
\caption{Top panel: DMX timeseries for Model E, where the constant and fixed linear trend components are not shown. Bottom panel: DMX timeseries minus the weighted average divided by the uncertainty on each measurement. All values \added{appear uncorrelated and}  fall within \replaced{1.4}{2.5}, i.e., all measurements are within \replaced{1.4}{2.5}$\sigma$ of the linear trend, suggesting there are no additional variations beyond the linear trend being measured by DMX.}


\end{figure}

\begin{itemize}
\item Model A, DMX: This is the same model as in NG15 using only DMX to model DM. For posterity we refit the timing model with \texttt{PINT} but as expected no significant parameter changes were found. Model A is our reference model against which we compare all of the other models.
\item Model B, Linear trend plus fixed solar wind density: We removed DMX, fit the constant term, and added and fit a linear trend term in DM, i.e., $\DM(t) = \DM_0 + (d\DM/dt)(t-t_0)$ where we set the epoch $t_0$ to be the same as the reference epoch for spin period and the astrometric terms. The parameter DM1 represents the quantity $d\DM/dt$. Not only do we find a significant linear trend in DM, we find a significant measurement of parallax inconsistent with the Model A upper limit; note that this inconsistency is unsurprising as the Lutz-Kelker bias will systematically overestimate the parallax even for the $\approx 4\sigma$ measurement in Model B \citep{lk73,vlm10}; applying a prior requiring non-negative parallaxes might bring the Model A value to within consistency as well. The component of the proper motion in ecliptic latitude also increases but not beyond the formal uncertainties on the parameter.
\item Model C, Quadratic trend plus fixed solar wind density: Similar to Model B but where we added the quadratic term in the polynomial Taylor expansion for the DM,
\be 
\frac{1}{2} \frac{d^2\DM}{dt^2} (t-t_0)^2,
\ee
to the equation. The parameter DM2 represents the quantity $d^2\DM/dt^2$.
\item \added{Model D, Linear trend plus fitted solar wind density: Similar to Model B but where we allowed the solar wind density to be a freely fit constant in the model.
}
\item Model D, Linear trend plus \replaced{varying solar wind density: To test whether we could improve our modeling of the solar wind to include time variability, we used a piecewise solar-wind electron density model (``SWX'' in \texttt{PINT}; \citealt{PINT2}) to model changes. Sixteen parameters were introduced to cover the timespan, where the starts and ends were defined to be halfway between each solar conjunction (minimum solar elongation) rather than arbitrarily set by the first observation.}{fitted constant solar wind density: To test our sensitivity to the solar wind, we allowed the solar wind density to be a free parameter. \texttt{PINT} also allows for the modeling of time variability with a piecewise solar-wind electron density model (``SWX'' in \texttt{PINT}; \citealt{PINT2}) but we found each parameter per year to be consistent with a constant density, the model be less favored statistically, and therefore we do not comment on this model further. }

\item Model E, Fixed linear trend plus fixed solar wind density and additional DMX: We took the constant and linear terms from Model B, fixed them in a new timing model, and added in DMX to look for individual departures in DM from Model B. The fit DMX values are shown in Fig.~\ref{fig:ModelE}, with no significant deviations away from zero (DM values away from the linear trend plus solar wind) found.
\end{itemize}

\begin{deluxetable*}{|l|c|c|c|c|}
\tablecolumns{5}
\tablecaption{Timing Models for \PSR}
\tablehead{
\colhead{} & \colhead{Model A:} & \colhead{Model B:} & \colhead{Model C:} & \colhead{Model D:} \\
\colhead{} & \colhead{Original DMX} & \colhead{Linear + Fixed SW} & \colhead{Quadratic + Fixed SW} & \colhead{Linear + Fit SW}
}
\label{table:J1455}
\startdata
\hline
\multicolumn{5}{c}{Dispersion Measure Model Components}\\
\hline
Constant & Yes (Fixed) & Yes & Yes & Yes\\
Piecewise-constant (DMX) & Yes & No & No & No\\
Linear Trend & No & Yes & Yes  & Yes \\ 
Quadratic Trend & No & No & Yes & No\\
Solar Wind & No & Yes (Fixed\tablenotemark{a}) & Yes (Fixed\tablenotemark{a}) & Yes (Fit) \\
\hline
\multicolumn{5}{c}{Spin and Astrometric Parameters}\\
\hline
Spin frequency, $(\nu_s - 125.20024515073)$ (10$^{-13}$ Hz) & $-$0.9(4) & 3.5(4) & 5.6(7) & 3.9(4) \\
Spin frequency derivative, $\dot{\nu}_s$ (10$^{-16}$ Hz s$^{-1}$) & $-$3.80963(3) & $-$3.80959(6) & $-$3.80963(6) & $-$3.80960(6) \\
Ecliptic longitude, $\lambda_e$ ($^\circ$) & 231.34754032(3) & 231.34754036(1) & 231.34754036(1) & 231.34754036(1) \\
Ecliptic latitude, $\beta_e$ ($^\circ$) & $-$16.04479841(11) & $-$16.04479839(5) & $-$16.04479839(5) & $-$16.04479836(6) \\
Proper motion in ecliptic longitude, $\mu_{\lambda_e}$ (mas yr$^{-1}$) & 8.10(3) & 8.09(1) & 8.09(1) & 8.09(1) \\
Proper motion in ecliptic latitude, $\mu_{\beta_e}$ (mas yr$^{-1}$) & 0.33(10) & 0.45(4) & 0.44(4) & 0.45(4) \\
Parallax, $\varpi$ (mas) & 0.23(33) & 0.94(17) & 0.91(17) & 0.87(19) \\
\hline
\multicolumn{5}{c}{Binary Parameters}\\
\hline
Orbital period, $P_b$ (days) & 76.174567480(5) & 76.174567472(2) & 76.174567471(2) & 76.174567472(2) \\
Projected semimajor axis, $x$ (lt-s) & 32.3622112(2) & 32.3622108(1) & 32.3622108(1) & 32.3622108(1) \\
Orbital eccentricity, $e$ & 0.00016964(1) & 0.00016966(1) & 0.00016966(1) & 0.00016966(1) \\
Epoch of periastron, $T_0$ (MJD) & 56064.3671(8) & 56064.3679(4) & 56064.3678(4) & 56064.3679(4) \\
Longitude of periastron, $\omega$ ($^\circ$) & $-$136.543(4) & $-$136.540(2) & $-$136.540(2) & $-$136.539(2) \\
Derivative of $x$, $\dot{x}$ (10$^{-14}$ s s$^{-1}$) & $-$2.4(2) & $-$2.1(1) & $-$2.1(1) & $-$2.1(1) \\
\hline 
Profile frequency dependency parameter, FD1 & 0.0000084(8) & 0.0000086(8) & 0.0000087(8) & 0.0000086(8) \\
\hline 
Dispersion measure, DM (pc cm$^{-3}$) & 13.570116 & 13.5698(1) & 13.5699(1) & 13.5699(1) \\
DM linear term, DM1 (pc cm$^{-3}$ yr$^{-1}$) & $-$ & 0.000118(8) & 0.000134(11) & 0.000117(8) \\
DM quadratic term, DM2 (pc cm$^{-3}$ yr$^{-2}$) & $-$ & $-$ & $-$0.000008(4) & $-$ \\
\hline 
$\Delta$ Akaike Information Criterion from Model C & 149.8 & 2.9 & 0.0 & 4.3 \\
$\Delta$ Bayesian Information Criterion from Model B & 1276.6 & 0.0 & 4.4 & 8.7 \\
\enddata
\tablenotetext{a}{Solar wind density fixed to a value of \replaced{7.9}{4.0}~cm$^{-3}$ from \replaced{\citet{Madison+2019}}{\citet{tempo2}}.}
\tablecomments{Reference Epoch 56079.0, orbital model DD \citep{DDI,DDII}.}
\end{deluxetable*}

In Table~\ref{table:J1455}, we report both the difference in the Akaike Information Criterion (AIC, $2k - 2\ln \hat{\mathcal{L}}$, for number of parameters $k$ and maximum log-likelihood value $\ln \hat{\mathcal{L}}$) and Bayesian Information Criterion (BIC, $k\ln N - 2\ln \hat{\mathcal{L}}$, where $N$ is the number of data points), based on the likelihood as calculated with \texttt{PINT} and the number of parameters, between \replaced{Model B and the other models, with Model B being the lowest, i.e., most preferred for both criteria.}{the preferred model and the others. Model C is preferred by the AIC and Model B is preferred by the BIC.} We do not report Model E in the table given the rationale above that no significant deviations were found -- the AIC and BIC values were unsurprisingly higher as it is still DMX-like. The fact that simpler models are preferred for both criteria is unsurprising given the number of DMX parameters (157) removed. \replaced{When we add the quadratic term in Model C, $\Delta\textrm{AIC} = 6.4$, and when converted into an evidence ratio implies Model B is 24 times more preferred than Model C. The limited significance of the quadratic term (DM2) supports this as well.}{When we add the quadratic term in Model C, the $\Delta\textrm{AIC}$ between Model B and C is 2.9, and when converted into an evidence ratio implies Model C is 4.3 times more preferred than Model B. When looking at the $\Delta\textrm{BIC}$, the difference is 4.4 but in favor of Model B, where the evidence ratio implies Model B is 9.0 times more preferred than Model C. Given the low levels of significance for adding a quadratic, and the $2\sigma$ measurement, we will largely focus on the linear trend model but will still comment on the implications of a higher-order trend.}

\added{We originally tried using the fixed solar wind density from \citet{Madison+2019} of 7.9 cm$^{-3}$ at 1 AU and found that Model B was preferred with both criteria. We found that using a density of 4.0 cm$^{-3}$ was preferred over 7.9 cm$^{-3}$ with $\Delta\textrm{AIC} = 11.7$ and $\Delta\textrm{BIC} = 8.8$. The fit solar wind density from Model D was $2.6 \pm 1.9$~cm~$^{-3}$ but the model was more disfavored in both the AIC and BIC values. When we fixed the solar wind density to this value, the AIC was 1.9 larger than for Model B while the BIC improved by 1.0. Our interpretation is that we are not particularly sensitive to the exact value of the solar wind density, as expected from our arguments in \S\ref{sec:heuristic} though lower values than in \citet{Madison+2019} have clear preference. One should not fix a value in the analysis without \textit{a priori} information, i.e., we could fix any other parameter in the model but do not, and so we discard models with this lower solar wind density and concentrate on those with a value of 4.0 cm$^{-3}$.}
  \added{We also tried a quadratic with a fit solar wind density but this was also disfavored for both criteria, and the tests of these sets of models demonstrate} \deleted{The linear trend model with a varying solar wind was even more disfavored, both in the AIC and BIC values. We show the SWX values (maximum DM from the solar wind as measured at conjunction) in Figure~\ref{fig:swx} and note that our values are consistent with both the solar wind density of 7.9~cm$^{-3}$ at 1 AU from \citet{Madison+2019} (translated into a DM of $4.0 \times 10^{-4}$~pc~cm$^{-3}$, see \S\ref{sec:solarwind}; dashed lines) but also zero (dotted lines), and again demonstrating } that this pulsar is not strongly sensitive to solar wind effects as expected. 
 \deleted{The additional model parameters are thus penalized by the AIC and BIC.}




We also tested various Bayesian Blocks \citep{Scargle+2013} algorithms iteratively in our fits to model DM as with DMX but with varying bin lengths, extending \texttt{Astropy}'s function \citep{astropy}. We applied the function to the TOAs multiplied by the radio-frequency squared (to scale proportionally to the DM). We tried using the function as provided by \texttt{Astropy}, modifying the algorithm to require window sizes between 2 and 90 days with at least 2 TOAs, and also trying both cases with a fixed solar wind electron density of \replaced{7.9}{4.0}~cm$^{-3}$ \deleted{\citep{Madison+2019}}. \replaced{The default Bayesian Blocks algorithm performed the best of these, but with a $\Delta\textrm{AIC} = 42.0$ worse than Model B ($\Delta\textrm{BIC} = 166.0$ worse). Therefore, while Bayesian Blocks may be useful in reducing the number of parameters in a piecewise DM model in general, here it is still strongly disfavored compared to the linear trend with a fixed solar wind density.}{The default Bayesian Blocks algorithm including the fixed 4.0 cm$^{-3}$ solar wind density performed the best of these, generating 19 DMX bins and resulting in $\Delta\textrm{AIC} = 68.2$ better than Model B (65.3 better than Model C) but a $\Delta\textrm{BIC} = 55.8$ worse than Model B. We see this as a viable alternative method to simplifying our DM modeling without sacrificing more complex time variability, and the method will be explored in future work (M. Thompson et al. in prep).}

In NANOGrav's standard timing analysis, we add or remove parameters using an $F$-test to check for statistical significance in particular orders. We opted to ignore this step in our particular analysis here given the overwhelming improvement between Model A and B, and the focus on this work in particular is the DM modeling alone. It is possible that given additional tweaks to the modeling, the difference between Model B and C could become much more favorable. Note that the only other radio-frequency-dependent parameter in the timing model is FD1, representing a timing delay equal to $\textrm{FD1} \times \ln(\nu)$, with $\nu$ in GHz. We see little change in the value of the parameter between models \added{compared to the uncertainties}, and if we add in the FD2 parameter (with time delay $\textrm{FD2} \times [\ln(\nu)]^2$) to Model B, we find that \replaced{both FD parameters become consistent with zero within the uncertainties}{FD1 is consistent with zero and FD2 is only a 1.4$\sigma$ significant parameter, and this model would be disallowed by NANOGrav's rules for adding in FD parameters \citep{NG15}}; the $\Delta$AIC from Model B is \replaced{28.9}{$-$0.1} \added{(essentially a tie)} but the $\Delta$BIC  \replaced{of 36.2}{is 7.1} \deleted{, which is even more disfavored than Model C}.

\section{Physical and Modeling Implications Considering Low-Order Polynomial Models for DM} 

In this section, we discuss the implications of replacing Model A (DMX) with Model B (linear trend) \added{and Model C (quadratic trend)} both on understanding the ISM and then with respect to GW sensitivity. We end with a discussion of the implications for DM modeling more broadly. \added{As discussed previously, we will focus more on the linear trend but include some discussion of a potential quadratic trend.}

\subsection{Implications for the ISM from the Linear/Quadratic Trends}
\label{sec:implicationsDM}

\replaced{While we constrain a value of the linear DM trend in Model B}{First, let us consider the linear DM trend in Model B. While we constrain a value for the trend}, we note that the measured value may indeed be contaminated by the stochastic variations we described in \S\ref{sec:turbulentDM}. We argued that such a \added{stochastic} component has an rms amplitude of approximately twice the median DM uncertainty \added{($2.5 \times 10^{-4}$~pc~cm$^{-3}$)} in \S\ref{sec:predictedDMvar}.
\added{A stochastic component will induce a DM gradient probabilistically, and assuming a Kolmogorov spectrum and frequency of 1300 MHz,} we therefore can calculate the rms \replaced{DM}{of this} gradient across the timespan $\tau = 15.7$~yr as \citep{DMt}
\be 
\sigma_{d\DM/dt} \approx \frac{\sigma_{\DM}(\tau)}{\tau} \approx 3.3 \times 10^{-5}~\mathrm{pc~cm^{-3}~yr^{-1}},
\ee
and so a deterministically-caused linear DM term might vary around the value we report by $\sim$30\%. \added{If instead we assumed the other possibility where $\alpha = 1.82$, then $\sigma_{d\DM/dt} \approx 8.0 \times 10^{-5}~\mathrm{pc~cm^{-3}~yr^{-1}}$, and so the linear trend we observe is of order the rms expectation, i.e., consistent within the range of expectations from stochasticity.}

Regardless, the value of the linear trend in Model B is an order of magnitude higher than the fiducial value \added{of $10^{-5}$~pc~cm$^{-3}$~yr$^{-1}$} presented in \S\ref{sec:linearDM}\added{, implying that the relative parallel velocity and/or the electron density must be higher than the fiducial values listed previously.}. We know of no known 3D velocity measurement for the pulsar \added{and so cannot directly estimate the parallel velocity}; the perpendicular component of the velocity as derived from the proper motion and parallax-inferred distance is $34 \pm 9$~km/s. This perpendicular velocity is consistent with the interstellar scintillation velocity estimator, $V_{\rm ISS} = 46 \pm 17$~km/s, calculated from \citet{cr98} for a uniform Kolmogorov medium,
\ba 
V_{\rm ISS} & = & A_{\rm ISS} \frac{\sqrt{D \Dnud}}{\nu \Dtd} \nonumber \\
& = & 2.53 \times 10^4~\mathrm{km/s}~ \frac{\displaystyle{\left(\frac{D}{\mathrm{kpc}}\right)^{1/2}\left(\frac{\Dnud}{\mathrm{MHz}}\right)^{1/2}}}{\displaystyle{\left(\frac{\nu}{\mathrm{GHz}}\right)\left(\frac{\Dtd}{\mathrm{s}}\right)}},
\ea
where $A_{\rm ISS}$ is a coefficient that depends on the physics and geometry of the intervening medium, and again we use the distance determined in Model B. Note that the increasing trend in DM does not inform which direction the pulsar is moving -- the density could increase due to an increasing line-of-sight distance but could also increase if the pulsar were moving towards us and ionizing the material in front of it \citep{DMt}. Even if the parallel component of the pulsar velocity reaches $\sim 100$~km/s, \added{in line with MSP velocity dispersions \citep{Matthews+2016},} then the local electron density at the pulsar needs to be $\sim 1$~cm$^{-3}$. The range of electron-density values in the Milky Way spans many orders of magnitude \added{($\sim0.1-10^4$~cm$^{-3}$)} but does include this value \citep{Draine}. \added{Therefore, realistic values for both the pulsar velocity and the electron density can produce the linear trend observed.}


\added{We will briefly comment on the Model C polynomial terms. As the DM polynomial is described as a Taylor-expansion around the reference epoch $t_{\rm ref}$, MJD 56079, the implied $\delta \mathrm{DM}$ from the quadratic term DM2 is
\be 
\delta \mathrm{DM}_{\rm quadratic} = \frac{1}{2} \left[\mathrm{DM2}\right] \left(t - t_{\rm ref}\right)^2 = -1.2 \times 10^{-7}~\mathrm{pc~cm}^{-3}.
\ee
Therefore, the quadratic term primarily serves to subtly alter the shape of the resultant polynomial fit rather than the endpoints; the implied $\delta\mathrm{DM}$ for the linear term is $\approx 0.002$~pc~cm$^{-3}$, which is apparent visually in Figure~\ref{fig:DMA}.

The quadratic component is likely not from the parallel motion of the pulsar. Following Eq.~\ref{eq:dDMdt} and its evaluation, we can write the second derivative of DM due to the parallel motion as $d^2 \DM/dt^2 \approx 10^{-5}~v_{100} \dot{n}_{e,0.1}$~pc~cm$^{-3}$~yr$^{-2}$, where the derivative of the electron density in units of 0.1~cm$^{-3}$~yr$^{-1}$, which would be an unsustainable decrease starting from a density of 1~cm$^{-3}$ over the 15.7-yr observing timespan. For a fiducial value of $C_n^2 \sim 10^{-3.5}~\mathrm{m}^{-20/3}$, the rms $n_e$ divided by the mean $n_e$ is approximately $10^{-4}$ \citep{cwb85}, implying comparatively small changes in $n_e$. The longer scintillation timescale implies a smaller value of the scattering measure, the integrated $C_n^2$ along the line of sight \citep{NE2001}, but the estimate remains similar and we do not expect such large changes in $n_e$ over the timespan we observe the pulsar moving. The distance change due to transverse motion is too small to change the DM by these amounts \citep{DMt}. The simplest local-density explanation is that if the pulsar parallel motion causes the linear trend, then the perpendicular motion causes the line of sight to pass through slightly less material, dropping the DM over time. However, we cannot confirm this, and instead if some or all of the observed DM we see is due to stochastic variations, then a quadratic term will naturally arise as a result. Further monitoring of the pulsar DM over time will help discriminate between possibilities.
}


\subsection{Implications for Sensitivity Toward Gravitational Waves and Timing Models Generally}

\label{sec:implicationsGW}

\begin{figure}[h!]
\hspace{-0.25in}
\includegraphics[width=0.53\textwidth]{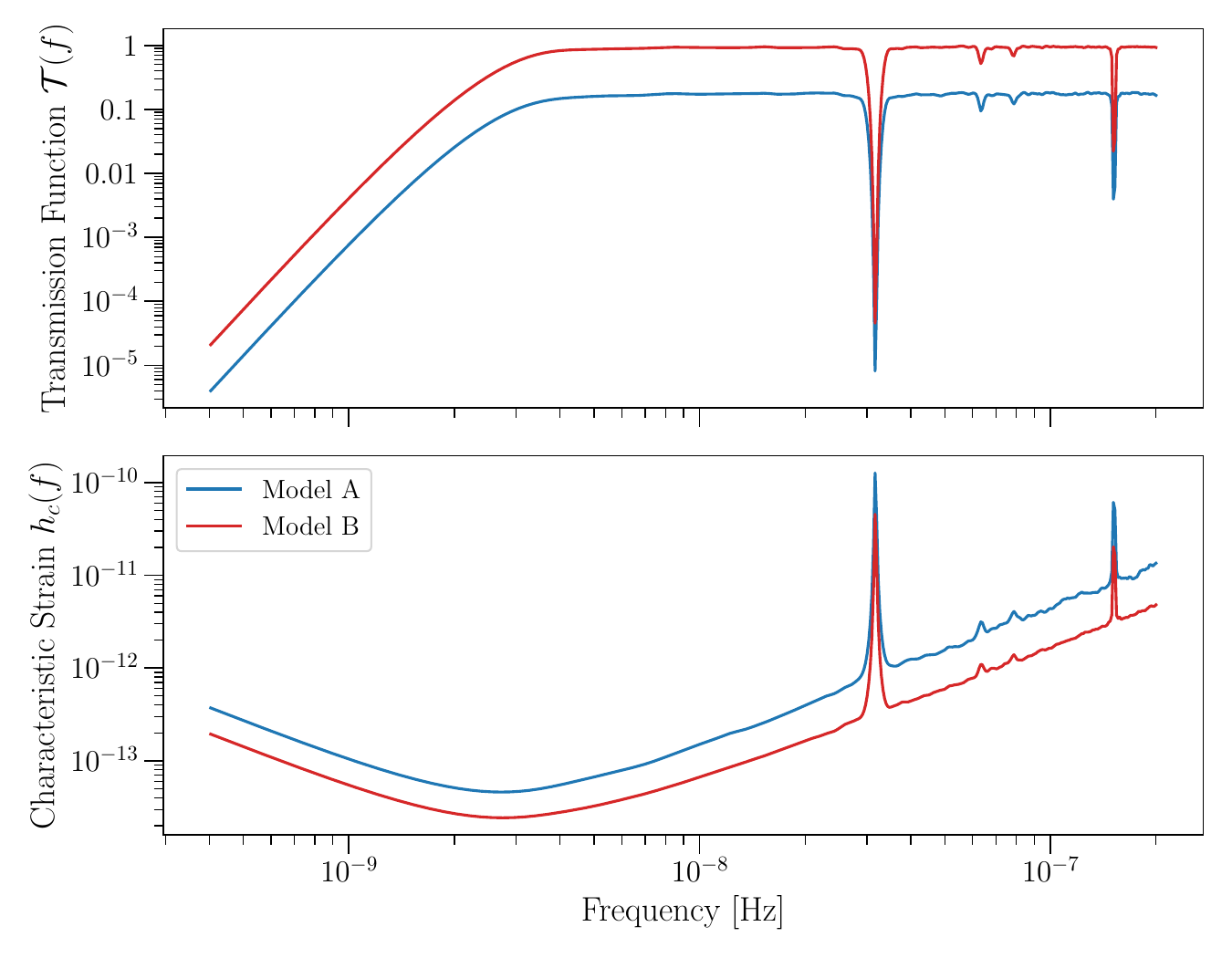}
\label{fig:sensitivity}
\caption{Transmission function (top) between the two different DM models, Model A (lower blue curve) and Model B (higher red curve), showing Model B recovers full transmission of power across frequencies. This recovery translates into an improvement in GW sensitivity (bottom) across all frequencies.}
\end{figure}

Using \texttt{hasasia} \citep{hasasia}, we computed transmission functions $\mathcal{T}(f)$ and GW sensitivity curves for Model A and Model B, shown in Fig.~\ref{fig:sensitivity}. \added{The Model C curve is very similar (though not identical) to the Model B curve on this scale and so we do not plot it.} The transmission function describes the amount of power removed from the TOAs as a result of fitting a specific timing model to the data. As shown in \citet{Hazboun+2019}, fitting for DMX results in a significant loss of power across GW frequency, and we see here that removing DMX and fitting the linear DM restores nearly all power ($\mathcal{T} \to 1$) at high frequencies. This restoration of power improves the GW sensitivity as shown in the bottom panel of Fig.~\ref{fig:sensitivity}, with the Model B \added{(and therefore also Model C)} curve below the Model A curve at all frequencies.

\begin{figure}[h!]
\centering
\includegraphics[width=0.48\textwidth]{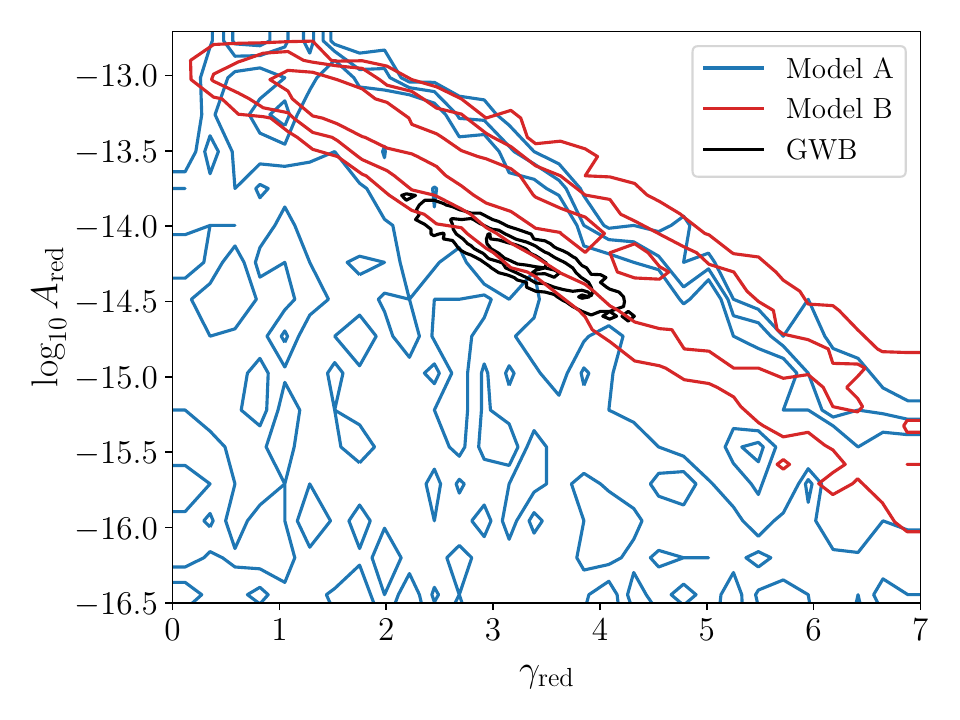}
\label{fig:posteriors}
\caption{Posteriors for red noise in Model A (blue), Model B (red), and the gravitational-wave background (GWB, black) reported in \cite{NG15GWB}. The amplitude $A_{\rm red}$ is in dimensionless GW strain units.}
\end{figure}

Without the peculiar absorption of power by the DMX model, while red noise was not significant in Model A, Model B shows significant red noise as defined by NANOGrav's standard criterion that the Savage-Dickey Bayes factor is greater than 100. We show the posteriors for the red noise for both models in Fig.~\ref{fig:posteriors}. \added{As before, the Model C posteriors overlap almost entirely with the Model B curves so we do not plot them.} \replaced{In addition, we overplotted}{We did overplot} the posteriors for the common \replaced{gravitational-wave background (GWB)}{GWB} signal as reported in \citet{NG15GWB}. The red noise is modeled with a power-law spectrum with amplitude $A_{\rm red}$, here shown in units of the logarithm of GW strain as performed by \texttt{enterprise} \citep{enterprise}. The spectral index is $\gamma_{\rm red}$, where the power spectrum is $S(f) \propto f^{-\gamma_{\rm red}}$ for GW frequency $f$. We see that the Model A posteriors are unconstrained, providing an upper limit in amplitude $A_{\rm red}$ as a function of spectral index $\gamma_{\rm red}$, whereas in Model B the posteriors are broad across the prior range but at least constrained to have a non-zero amplitude. The posteriors are consistent with the GWB posteriors. This implies that the replacement of DMX, a radio-frequency-dependent model, (re-)introduces both a timing parallax signal and red noise signal in the fitting process, both frequency-independent timing perturbations, and therefore modeling DMX will cause both signatures to cancel out. \added{Therefore, we find that at least in this one pulsar that DMX is absorbing our GW signal of interest. Achromatic red noise absorption by DMX was shown in \citet{Hazboun+2019} and we know that it is a systematic bias in our GWB parameter estimation \citep{NG15GWB}. We recommend considering more simplified DM models, where appropriate, to minimize these biases.}

\PSR\ is also observed by the European Pulsar Timing Array (EPTA; \citealt{EPTA}) and the MeerKAT Pulsar Timing Array (MPTA; \citealt{MPTA}) as part of their PTAs. In its second data release, the EPTA models DM up to the quadratic term across all of its pulsars, fixes the solar wind density, and then models remaining variations in the Fourier domain with a power-law spectrum \citep{EPTADR2Noise}. Certain specific pulsars include additional components not relevant for \PSR. They do not find any significant trends or remaining variations for this pulsar based on their independent 15.7 years of data given the uncertainties. However, our Model B parallax value is consistent \replaced{with}{within $\sim 2\sigma$ of} their value of $1.3 \pm 0.1$~mas. The EPTA also measures red noise in this pulsar, consistent with the Model B contours in Figure~\ref{fig:posteriors} for the full version of its data set. These similarities between data sets lend support to the fact that in the NANOGrav data set, DMX is absorbing the parallax signal, which in turn alters the measured achromatic red noise properties \citep{PhilMemo}. Similar to the EPTA, the MPTA also models DM up to the quadratic term and then similarly models remaining variations in the Fourier domain with a power-law spectrum \citep[see also][]{MPTA2}, and finds a consistent parallax value of $1.1 \pm 0.3$~mas \citep{MPTAastrometric}.

\added{
We note that between Model A and Models B or C, many parameter uncertainties have improved. The inconsistency of the spin frequency measurements is unsurprising given that the achromatic red noise is strongly covariant and has changed between runs. The astrometric parameter uncertainties have improved by a factor of 2--3 as have a number of the binary parameters. This is again unsurprising as DMX is absorbing power in the timing fit away from these other parameters, increasing the uncertainties; the reverse is happening when we remove DMX. The use of simplified DM modeling instead of DMX may therefore improve parameter estimation for pulsar studies more generally \citep[see e.g.,][]{Fonseca+2021}.
}

\added{A recent study of the interplay between DM methods and absorption of achromatic red noise was performed by \citet{Iraci+2024} using simulations. They found that when the red noise amplitude was large, DMX was sufficiently accurate in recovering both the injected DM variations and red noise. However, when red noise is not included in the fit or when the red noise amplitude is low, DMX can absorb this power. Both conditions are true for NANOGrav's timing of this pulsar. In our standard practices, NANOGrav removes red noise altogether when constructing timing models if the significance is too low, though individual red noise components are always added in the GW modeling \citep{NG15GWB}. Regardless, the red noise amplitude is still low compared to the white noise and the spectral properties of the red noise are poorly constrained(see Figure~\ref{fig:posteriors}). Therefore, our results for \PSR\ appear consistent with the findings by \citet{Iraci+2024}.

}

\subsection{Implications for DM Modeling in Precision Timing}

This simplified DM modeling is physically motivated by the long scintillation timescale and uncertainties on the DM -- it will not apply to most pulsars in PTAs currently observed \citep[e.g.,][]{Levin+2016,Turner+2021}. \added{Briefly exploring the DM variations in our sample further, when we fit a linear trend directly to the NANOGrav 15-year data set DMX timeseries (rather than refit in the timing model), we see that 13/68 pulsars have a reduced chi-squared less than three but only three of those have timespans longer than five years: PSRs~J1453+1902 (7 years), J1455$-$3330 (15.7 years), and J2214+3000 (8.4 years). PSR~J1453+1902 has no values $\gtrsim 3\sigma$ away from the fit and PSR J2214+3000 has only two $\sim 4\sigma$ departures away. These other two pulsars do not have scintillation timescales measured but the NE2001-derived values at 1500 MHz are 26 and 41 minutes, respectively \citep{NE2001,Turner+2021}, and as such simpler models for variations covering intermediate timescales between the observing cadence (unlike DMX) and the observing timespan (unlike this work) may be preferable. We find that additional variations on short and long timescales appear for more pulsars when we increase our reduced chi-squared threshold. Analyses separating deterministic and stochastic trends in NANOGrav's 15-year data set is an ongoing effort (A. K. Sreekumar et al. in prep.) and may inform the DM modeling in future data sets.
}

As surveys find more pulsars, those will tend to be more distant than the current sample with scintles smaller than the current average. Nonetheless, even with the current sample, the scintillation timescale should be considered in terms of DMX binning \citep[][]{NG9DM} (included in a standard implementation or one using Bayesian Blocks) or the resolvable scales for Gaussian Process methods as it provides prior information as to the scale of DM fluctuations caused by turbulence. Such a prior may miss rapid discrete changes in DM (true changes or apparent radio-frequency-dependent changes in the arrival times), which should be modeled separately regardless given the available observables, e.g., high-resolution and broadband dynamic spectra. Cyclic spectroscopy \citep{Demorest2011,WDvS2013,Dolch+2021,Turner+2024}, e.g., with a real-time backend such as the one in design at the Green Bank Observatory, coupled with the latest and upcoming generation of ultrawideband receivers will provide information which may help in uncoupling the sources of such discrete events.

\begin{acknowledgments}

{\it Author Contributions.} 
M.T.L. performed the majority of the analyses presented in this work and wrote the manuscript. D.L.K. performed the Bayesian Blocks analysis. P.T.B., J.S.H., and J.S. performed advanced per-pulsar noise modeling on earlier NANOGrav data sets which uncovered the first preference for a linear DM trend. \added{D.J.N. and O.A.C. generated the cleaned flux density measurements.} All other authors along with M.T.L. created the curated NANOGrav 15-year data set, and specific contributions are summarized in \citet{NG15}. T.T.P. led the long-track Shapiro delay observations used in our dynamic spectrum analysis.

{\it Acknowledgments.}
\added{We thank the anonymous referee for suggestions which substantially improved this work.}
We acknowledge support received from NSF Physics Frontiers Center award number 2020265, which supports the NANOGrav project.
P.R.B. is supported by the Science and Technology Facilities Council, grant number ST/W000946/1.
Pulsar research at UBC is supported by an NSERC Discovery Grant and by CIFAR. \linebreak

K.C. is supported by a UBC Four Year Fellowship (6456).
M.E.D. acknowledges support from the Naval Research Laboratory by NASA under contract S-15633Y.
T.D. and M.T.L. are supported by an NSF Astronomy and Astrophysics Grant (AAG) award number 2009468.
E.C.F. is supported by NASA under award number 80GSFC21M0002.
G.E.F. is supported by NSF award PHY-2011772.
D.R.L. and M.A.M. are supported by NSF \#1458952.
M.A.M. is supported by NSF \#2009425.
The Dunlap Institute is funded by an endowment established by the David Dunlap family and the University of Toronto.
T.T.P. acknowledges support from the Extragalactic Astrophysics Research Group at E\"{o}tv\"{o}s Lor\'{a}nd University, funded by the E\"{o}tv\"{o}s Lor\'{a}nd Research Network (ELKH), which was used during the development of this research.
N.S.P. was supported by the Vanderbilt Initiative in Data Intensive Astrophysics (VIDA) Fellowship.
H.A.R. is supported by NSF Partnerships for Research and Education in Physics (PREP) award No. 2216793.
S.M.R. and I.H.S. are CIFAR Fellows.
Portions of this work performed at NRL were supported by ONR 6.1 basic research funding.
J.S. is supported by an NSF Astronomy and Astrophysics Postdoctoral Fellowship under award AST-2202388, and acknowledges previous support by the NSF under award 1847938.

\end{acknowledgments}

\software{\texttt{Astropy} \citep{astropy}, \texttt{PINT} \citep{PINT,PINT2}, \texttt{enterprise} \citep{enterprise}, \texttt{hasasia} \citep{hasasia}}

\appendix
\added{
\section{Flux Density Measurements for \PSR}
\label{sec:appendix}

We present the timeseries of flux density measurements for our 820 and 1500~MHz receivers from the NANOGrav 15-year data set. The flux density calculations follow the procedures described in the NANOGrav 12.5-year data set release \citep{NG12p5} and are briefly outlined here. Only GUPPI data were used as the narrower bandwidth of GASP did not allow for more robust cross-checks. Absolute flux calibration was performed as described in \S\ref{sec:obs} above and further in \citet{NG12p5} and \citet{NG15}. For each epoch and frequency band, flux density values were compared with those at nearby frequencies and were flagged according to a metric involving the ratio of the flux density $S$ divided by the S/N along with corrections to compare the radiometer noise, in total $(S/[\mathrm{S/N}])\sqrt{2Bt}$ where $B$ is the channel bandwidth and $t$ was the integration time. If this metric was too low ($<0.75$) or high ($>1.75$) compared to the median for groups of 50 MHz and 100 MHz for the 820 and 1500 MHz receivers, respectively, then that measurement was flagged. Epochs with more than five flagged points were removed entirely.

We modeled the flux density measurements as a power law and performed a linear fit in log-log space to values at each epoch and receiver. Figure~\ref{fig:flux} shows the flux density timeseries. Most epochs for the GBT 13A--446 Shapiro delay observing campaign were flagged as a result of the procedure above, including the two long tracks mentioned in this work and specifically MJD 56645 from which we computed the dynamic spectrum in Figure~\ref{fig:dynspec} and calculated scintillation parameters. To place these measurements on the plot, we took the dynamic spectrum and computed the median and the range encompassing the inner 95\% of the values to show the spread of values.

 We also show the structure function of each timeseries, logarithmically binned in lag, ignoring the range of measurements from MJD 56645. The flat shape in each indicates no strong correlations present in either timeseries \citep{sc90}. We also calculated the modulation indices for each timeseries, defined as the rms divided by the mean flux density, 
and obtained values of $0.61 \pm 0.02$ and $0.65 \pm 0.01$ for the 820 and 1500 MHz timeseries, respectively. Correcting for noise fluctuations (removing in quadrature the median fractional uncertainty on the flux density values; \citealt{bgr99}) does not change our reported values within significant digits. The lack of correlations and the moderate modulation indices both support the flux density variations being driven by diffraction rather than refraction.

\begin{figure}[t]
\centering
\hspace{-0.2in}
\includegraphics[width=0.7\textwidth]{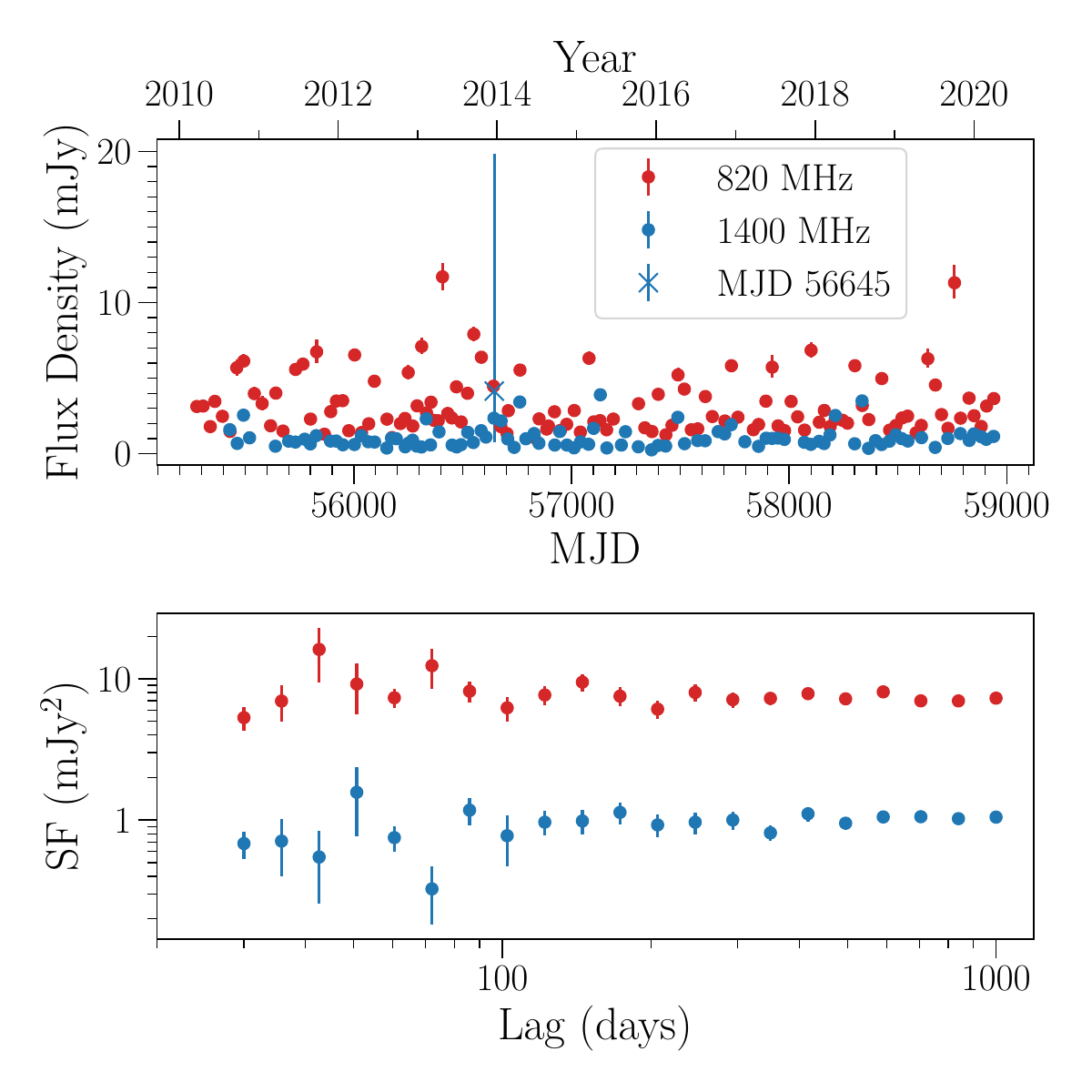}
\label{fig:flux}
\caption{Top panel: Flux density measurements for profiles observed at each of the two bands. Our long track observation is shown with the cross, where the error bar for that point denotes the inner 95\% range of values rather than the standard uncertainty. Bottom panel: Flux density structure function for each of the two timeseries, showing no strong correlations present.
}
\end{figure}
}

\end{document}